\documentclass[12pt]{article}
\usepackage{times}
\usepackage[dvips]{graphicx}

\title{Self-organization of oscillation in an epidemic model for COVID-19}

\author{
Takashi Odagaki$^\ast$\\
\\
\normalsize{Kyushu University}\\
\normalsize{Nishiku, Fukuoka 819-0395, Japan}\\
\normalsize{and}\\
\normalsize{Research Institute for Science Education, Inc.}\\
\normalsize{Kitaku, Kyoto 603-8346, Japan}\\
\normalsize{$^\ast$Corresondence to: Research Institute for Science Education, Inc.,}\\
\normalsize{Kitaku, Kyoto 603-8346, Japan}\\
\normalsize{{\it Email address}: t.odagaki@kb4.so-net.ne.jp}\\
}

\date{\today}

\begin{document} 

\baselineskip24pt

\maketitle 


\begin{abstract}
On the basis of a compartment model, the epidemic curve is investigated when the net rate $\lambda$ of
change of the number of infected individuals $I$  is given by an ellipse in the $\lambda$-$I$ plane
which is supported in $[I_{\ell}, I_h]$.
With $a \equiv (I_h - I_{\ell})/(I_h + I_{\ell})$, it is shown that (1) when $a < 1$ or $I_{\ell} >0$,
oscillation of the infection curve is self-organized and the period of the oscillation is
in proportion to the ratio of the difference $ (I_h - I_{\ell})$ and
the geometric mean $\sqrt{I_h I_{\ell}}$ of $I_h$ and $I_{\ell}$,
(2) when $a = 1$, the infection curve shows a critical behavior where it decays obeying a power
law function with exponent $-2$ in the long time limit after a peak, and
(3)  when $a > 1$, the infection curve decays exponentially in the long time limit
after a peak.
The present result indicates that the pandemic can be controlled by a measure which
makes $I_{\ell} < 0$.
\end{abstract}


\section{\label{sec:level1}Introduction}

Since the first outbreak in China in November 2019, COVID-19 has been spreading in
all continents including Antarctica.
According to a recent analysis of infection status of 186 countries \cite{JHU, oda-suda},
the time dependence of the daily confirmed new cases in more than 80 countries
show oscillations whose periods range from one to five months depending on the country.
The period of the oscillation is much shorter than that of 
Spanish flu in 1918$\sim$1919 which is the result of the mutation of virus,
and it is an open question why the infection curve of COVID-19
shows oscillation in some countries.

There have been several compartmental models which explain epidemic
oscillations \cite{hethcote, earn, zhang}.
The simplest idea to explain the oscillation is to introduce a sinusoidal time dependence
of parameters of the model.
Recently, Greer et al \cite{greer} introduced a dynamical model with time-varying
births and deaths which shows oscillations of epidemics. 

Since the infection curve of COVID-19 shows different features depending on the country,
the infection curve must have a strong relation to
the government policy, and the conventional approach may not be appropriate to COVID-19.
In fact, different measures have been employed in each country by its government and citizens
have been restricting the social contact among them, both of which depend on the infection status.
Therefore, parameters including transmission coefficient of the virus can be considered
to be a function of the infection status, and the non-linear effects due to
this dependence must be clarified.

In this paper, I introduce a compartment model in which the net rate $\lambda$ of
change of the number of infected individuals $I$ is a function of $I$ and
the function is given by an ellipse in the $\lambda$-$I$ plane
which is supported in $[I_{\ell}, I_h]$. Here, $I_h$ is the upper limit of
the number of infected individuals
above which the government does not allow, and $I_{\ell}$ is the lowest
value below which the government will lift measures.
I show that an oscillatory infection curve can be self-organized when $I_{\ell} >0$
and that the period is determined by the ratio of the difference $I_h - I_{\ell}$ and
the geometric mean $\sqrt{I_h I_{\ell}}$ of $I_h$ and $I_{\ell}$. I also show that when $I_{\ell} = 0$ the infection curve
in the long time limit after a single peak decays following a power law function with exponent -2 and
when $I_{\ell} < 0$ it decays exponentially in the long time limit.

\section{\label{Sec:2}Model country}
In most of compartmental models for epidemics, the number of infected individuals $I(t)$
is assumed to obey
\begin{equation}
\frac{d I(t)}{dt} = \lambda I(t).
\label{eq:1}
\end{equation}
The net rate of change $\lambda$ of the number of infected individuals is written generally as
\begin{equation}
\lambda = \beta\frac{S}{N} - \gamma - \alpha.
\label{eq:2}
\end{equation}
Here, $\beta$ and $\gamma$ are the transmission rate of virus from an infected individual
to a susceptible individual and a per capta rate for becoming a recovered non-infectious (including dead)
individual (R), respectively, and $S$ and $N$ are the number of
susceptible individuals and the total population.
In Eq.~(\ref{eq:2}), $\alpha$ is a model-dependent parameter representing different effect
of epidemics.
In the SIR model \cite{SIR}, it is assumed that no effects other than transmission and recovery
are considered and thus $\alpha =0$ is assumed.
The SEIR model \cite{SEIR} introduces a compartment of exposed individuals (E),
and if one sets $\alpha = (dE/dt)/I$, the basic equation of the SEIR model reduces
to Eq.~(\ref{eq:1}).

The SIQR model \cite{Hethcote-SIQR, oda-idm} separates quarantined patients (Q) as a compartment
in the population and $\alpha$ in Eq.~(\ref{eq:1}) is given by
the quarantine rate $q \equiv \Delta Q(t)/I(t)$ where $\Delta Q(t)$ is
the daily confirmed new cases \cite{oda-wavy}.
In the application of the SIQR model to COVID-19, it has been shown that 
\begin{equation}
\Delta Q(t) \propto I(t - \tau),
\label{eq:3}
\end{equation}
where $\tau$ is a typical value of the waiting time between the infection and quarantine
of an infected individual.
Therefore, the number of the daily confirmed new cases can be assumed to obey Eq.~(\ref{eq:1}) 
with the redefined time $t-\tau$. Since $\Delta Q (t)$ is treated as an explicit variable
instead of $I(t)$,
the SIQR model is relevant to COVID-19.

In this paper, I focus on the time evolution of $I(t)$ governed by Eq. (\ref{eq:1}) for COVID-19.
The transmission coefficient is determined by characteristics of the virus and by government policies
for lockdown measure and vaccination and by people's attitude for social distancing.
Medical treatment of infected individuals affects $\gamma$ and the government policy on PCR test
changes the quarantine rate.
The government policies are determined according to the infection status and therefore
the rate of change is considered to be a function of $I(t)$ in Eq. (\ref{eq:1}).

Here, I consider a model country in which $\lambda$ depends on $I$ through
\begin{equation}
\left( \frac{\lambda}{\lambda_0} \right)^2 + \left( \frac{I-I_0}{\Delta} \right)^2 = 1.
\label{eq:4}
\end{equation}
This implies that when $I$ becomes large, some policies are employed to reduce $\lambda$ to
the negative area so that $I(t)$ begins to decline
and when $I$ becomes small enough, then some measures are lifted
and $\lambda$ becomes positive again. In fact, the plots of $\lambda(t)$ against $I(t)$
in many countries show similar loops \cite{oda-suda}.
Note that $\lambda = 0$ corresponds to either a maximum or a minimum
of the number of infected individuals.

Figure 1 shows this dependence, namely $I_{h}\equiv I_0 + \Delta$ and 
$I_{\ell} \equiv I_0 - \Delta$ are the maximum and minimum
of the number of infected individuals set by the policy in the country.
When $I_ {\ell} < 0$, $\lambda$ in the region $I < 0$ is not relevant
since no infected individuals exist in this region.
\begin{figure*}[h]
\begin{center}
\includegraphics[width=7cm]{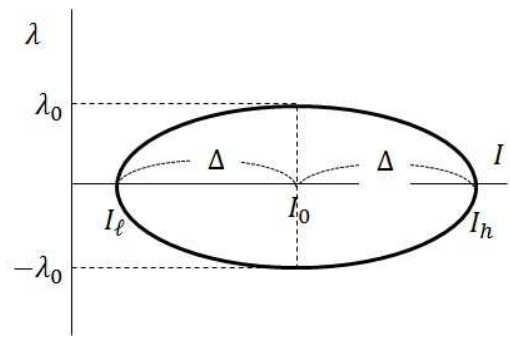}
\caption{\label{fig1}
The dependence of the net rate $\lambda$ on the number of infected individuals $I$ in a model country.
}
\end{center}
\end{figure*}

\section{\label{Sec:3} Infection curve and self-organization of oscillation}
In order to solve Eq.~(\ref{eq:1}) with Eq.~(\ref{eq:4}), I introduce a variable $x$ through
\begin{eqnarray}
\lambda &=& \lambda_0 \cos x ,\\
I - I_0 &=& \Delta \sin x 
\end{eqnarray}
and rewrite Eq.~(\ref{eq:1}) as
\begin{equation}
a \frac{d x}{d\tilde{t}} = 1 + a \sin x,
\label{dxdt}
\end{equation}
where $\tilde{t} \equiv \lambda_0 t$ is the time scaled by $\lambda_0^{-1}$ and
$a \equiv \Delta/I_0 \ge 0$ is a parameter of the model.
Equation (\ref{dxdt}) can be solved readily under the initial condition $I(t=0) = I_0$:
\begin{equation}
\tilde{t} = \left\{ \begin{array}{ll}
\frac{\displaystyle 2a}{\displaystyle \sqrt{1-a^2}} \left[ \arctan\frac{\displaystyle \tan(x/2)+a}{\displaystyle \sqrt{1-a^2}}
- \arctan\frac{\displaystyle a}{\displaystyle \sqrt{1-a^2}}\right] & \quad \mbox{when $a < 1$}, \nonumber\\
\frac{\displaystyle 2\tan(x/2)}{\displaystyle 1 + \tan(x/2)} &\quad \mbox{when $a = 1$}, \\
\frac{\displaystyle a}{\displaystyle \sqrt{a^2-1}} \left[ \ln \frac{\displaystyle \tan(x/2)+a-
\sqrt{a^2-1}}{\displaystyle \tan(x/2)+a+\sqrt{a^2-1}} - 
\ln \frac{\displaystyle a-\sqrt{a^2-1}}{\displaystyle a+\sqrt{a^2-1}}  \right]&\quad \mbox{when $a > 1$}. \nonumber
\end{array} \right.
\end{equation}
The infection curve is given in terms of $\tan(x/2)$ by
\begin{equation}
\frac{I(t)}{I_0} = 1 + \frac{a \tan(x/2)}{1 + \tan^2(x/2)}.
\end{equation}
The infection curves are shown for $a = 0.4, 0.6, 0.8, 1$ in Fig. 2(a) and
for $a = 1, 2, 4, 6$ in Fig. 2(b).
Therefore, the infection curve is a periodic function when $a < 1$
and a decaying function with a single peak when $a \ge 1$.
\begin{figure*}
\begin{center}
\includegraphics[width=6cm]{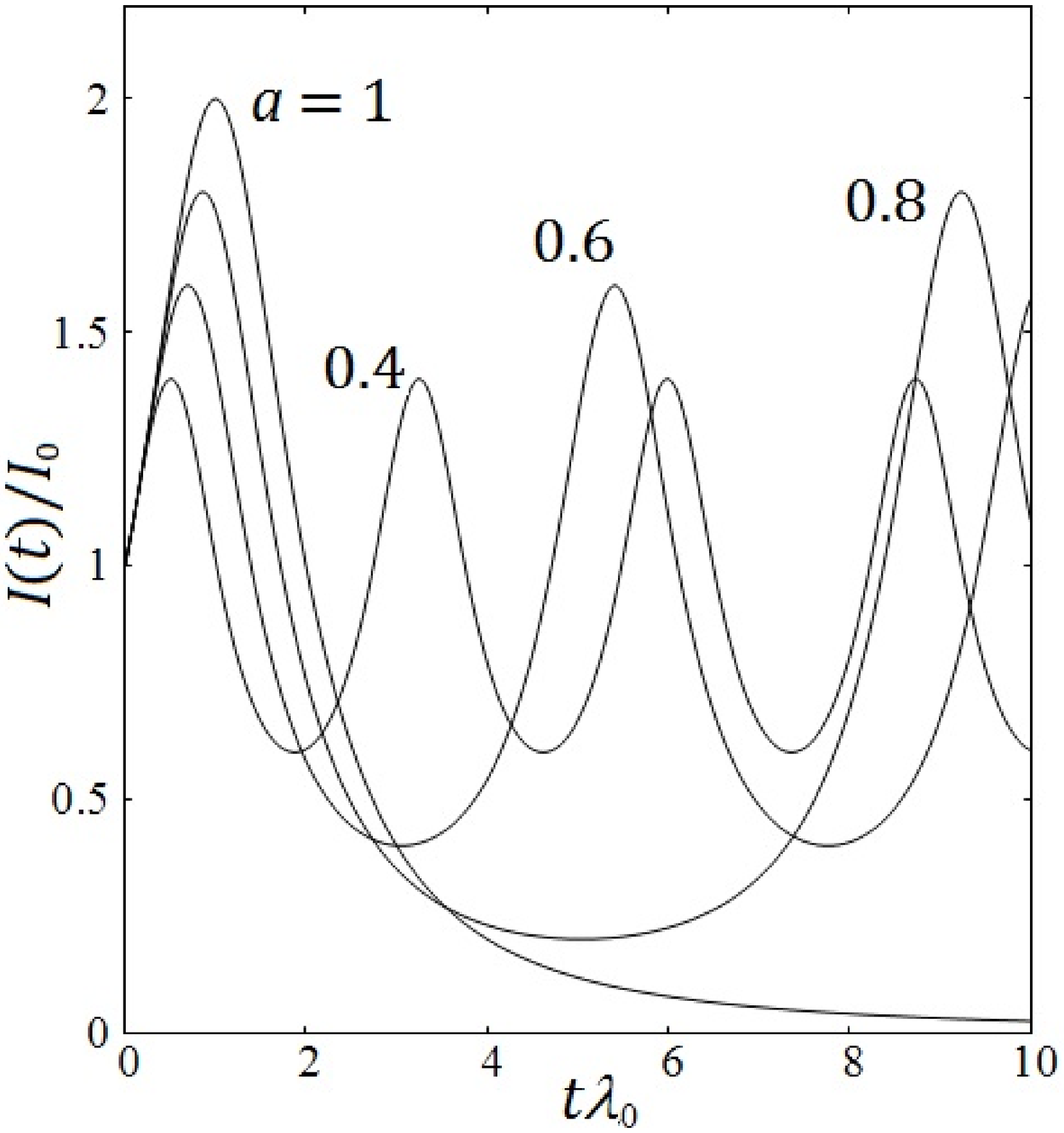}\hspace{1cm}
\includegraphics[width=6cm]{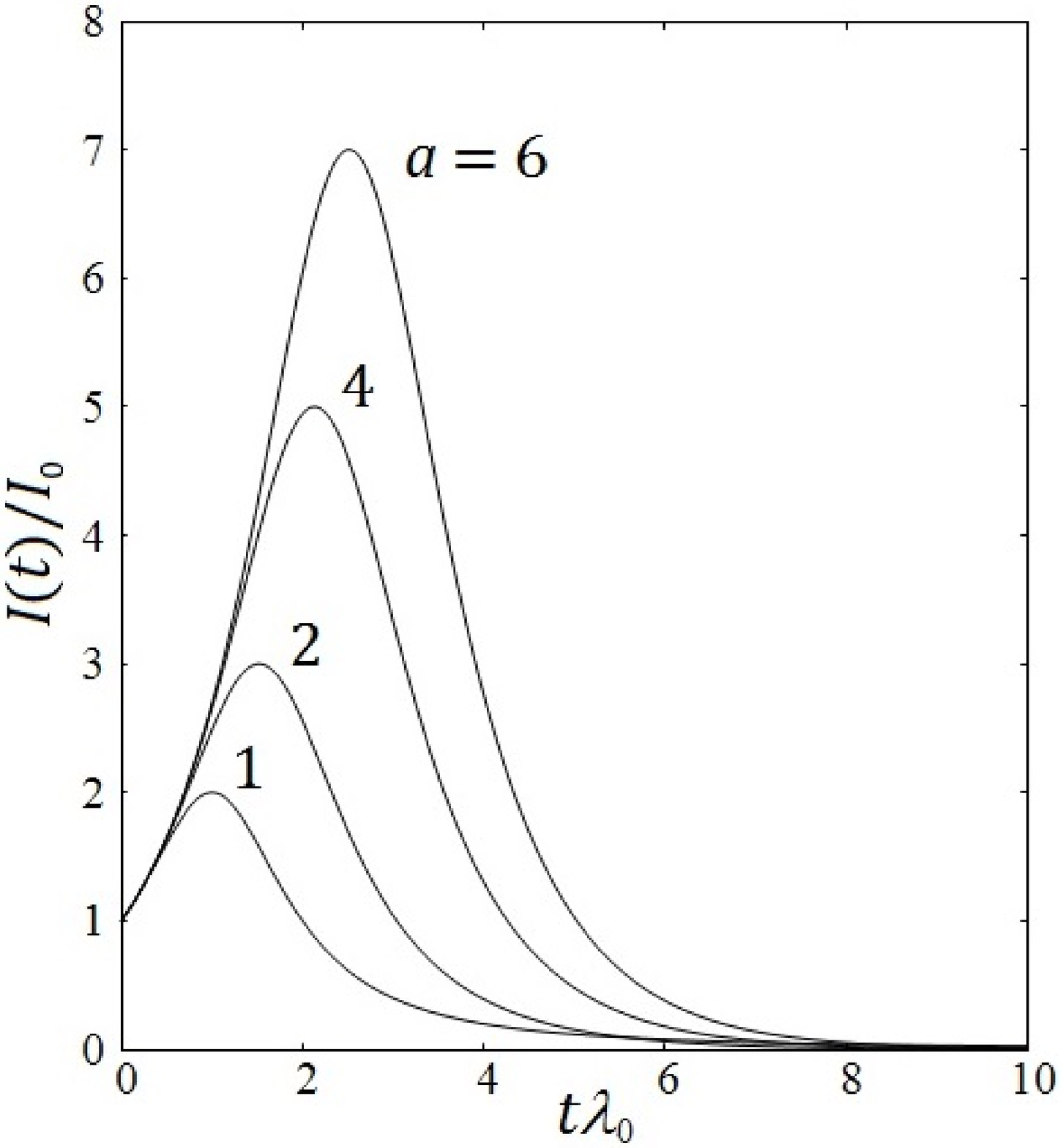}\\
(a)\hspace{6.5cm}(b)
\caption{\label{fig2}
The infection curve for the model country. (a) When $a<1$, a wavy infection curve is
self-organized. (b) When $a>1$ the infection curve is a decaying function
with a single peak. The infection curve for $a=1$ shown in both panels obeys
a power-law decay in the long time limit after a peak.
}
\end{center}
\end{figure*}

Characteristics of the infection curve are in order:\\
(1) When $a < 1$, the infection curve shows a self-organized oscillation
which can be characterized as follows:
\begin{enumerate}
\item The location of the peak $I_{\rm max}/I_0 = 1+a =I_h/I_0$ and the bottom
$I_{\rm min}/I_0 = 1- a = I_{\ell}/I_0$ are given by
\begin{eqnarray}
\tilde{t}_{\rm max} (n) &=& \frac{2a}{\sqrt{1-a^2}}\arctan\left[ 
\sqrt{\frac{1-a}{1+a}} +(n-1)\pi \right],\\
\tilde{t}_{\rm min} (n) &=& \frac{2a}{\sqrt{1-a^2}}\arctan \left[
\sqrt{\frac{1+a}{1-a}} + (n-1)\pi\right],
\end{eqnarray}
respectively, where $n = 1, 2, \dots$. 
\item Therefore, the period $T$ is given by
\begin{equation}
T\lambda_0 = \frac{2\pi a}{\sqrt{1-a^2}} 
= \frac{4\pi(I_h - I_{\ell})}{\sqrt{I_h I_{\ell}}}.
\end{equation}
Namely, the period is given by the ratio of a half of
the difference $\Delta= \frac{I_h - I_{\ell}}{2}$
and the geometrical mean $\sqrt{I_h I_{\ell}}$ of $I_h$ and $I_{\ell}$.
\end{enumerate}

\noindent
(2) When $a = 1$, the infection curve shows a peak, after which it decays to zero. It can be characterized
as follows:
\begin{enumerate}
\item The infection curve reaches its maximum $I_{\rm max}/I_0 = 2$ at $t\lambda_0 = 1$.
\item In the long time limit, it decays as $ t^{-2}$.
\end{enumerate}

\noindent
(3) When $a > 1$, the infection curve shows a peak, after which it decays to zero. It can be characterized
as follows:
\begin{enumerate}
\item The infection curve reaches its maximum $I_{\rm max}/I_0 = 1+a$ at
 $t\lambda_0 = \frac{a}{\sqrt{a^2-1}} \ln (a+\sqrt{a^2-1})$.
\item The infection curve returns to the initial state $I(t) = I_0$ at
$t\lambda_0 = \frac{a}{\sqrt{a^2-1}} \ln \frac{a+\sqrt{a^2-1}}{a-\sqrt{a^2-1}}$.
\item In the long time limit, the effective relaxation time defined by
$ \tau \equiv -\left(\frac{d\ln I}{dt}\right)^{-1}$ is given by $\tau\lambda_0 = \frac{a}{\sqrt{a^2-1}}$.
\end{enumerate}

Figure 3 shows the period for $a < 1$ and the relaxation time for $a > 1$
as functions of $a$.

\begin{figure*}
\begin{center}
\includegraphics[width=7cm]{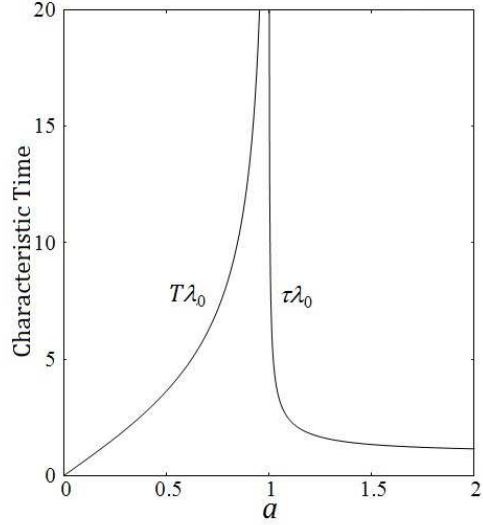}
\caption{\label{fig3}
The period when $ a <1$ and the relaxation time in the long time behavior when $a > 1$
are shown as functions of $a$.
}
\end{center}
\end{figure*}

\section{Discussion}
I have shown that oscillation of the infection curve can be self-organized in
the epidemic model described by an ordinary differential equation which exploits
the net rate of change Eq.~(\ref{eq:4}) depending on the number of infected individuals.
All countries employ their own policy which depends on the infection status of
the country
and the relation Eq.~(\ref{eq:4}) represents general trend of the policy.
Namely, when the number of infected individuals approaches the maximum number acceptable
in a country, a strong measure is introduced to make the net rate of
change $\lambda$ negative, and the measure will be lifted when the number of
infected individuals is considered to be small enough, which makes $\lambda >0$.
Therefore, the policy with $I_{\ell}>0$ itself is considered to be the origin
of the oscillation of the infection curve and the policy with $I_{\ell} < 0$
seems to have succeeded in controlling the pandemic \cite{oda-suda}.
 As an example, I show in Fig. 4(a) the time dependence
of daily confirmed new cases in Japan from April 5, 2020 to February 11, 2021 which consists
of three waves. Using $\lambda$ determined by fitting the data by 
piece-wise quadratic functions as shown by the solid curve \cite{oda-suda},
I show the correlation between $\lambda$
and $\Delta Q$ in Fig. 4(b).
\begin{figure*}
\begin{center}
\includegraphics[width=5cm]{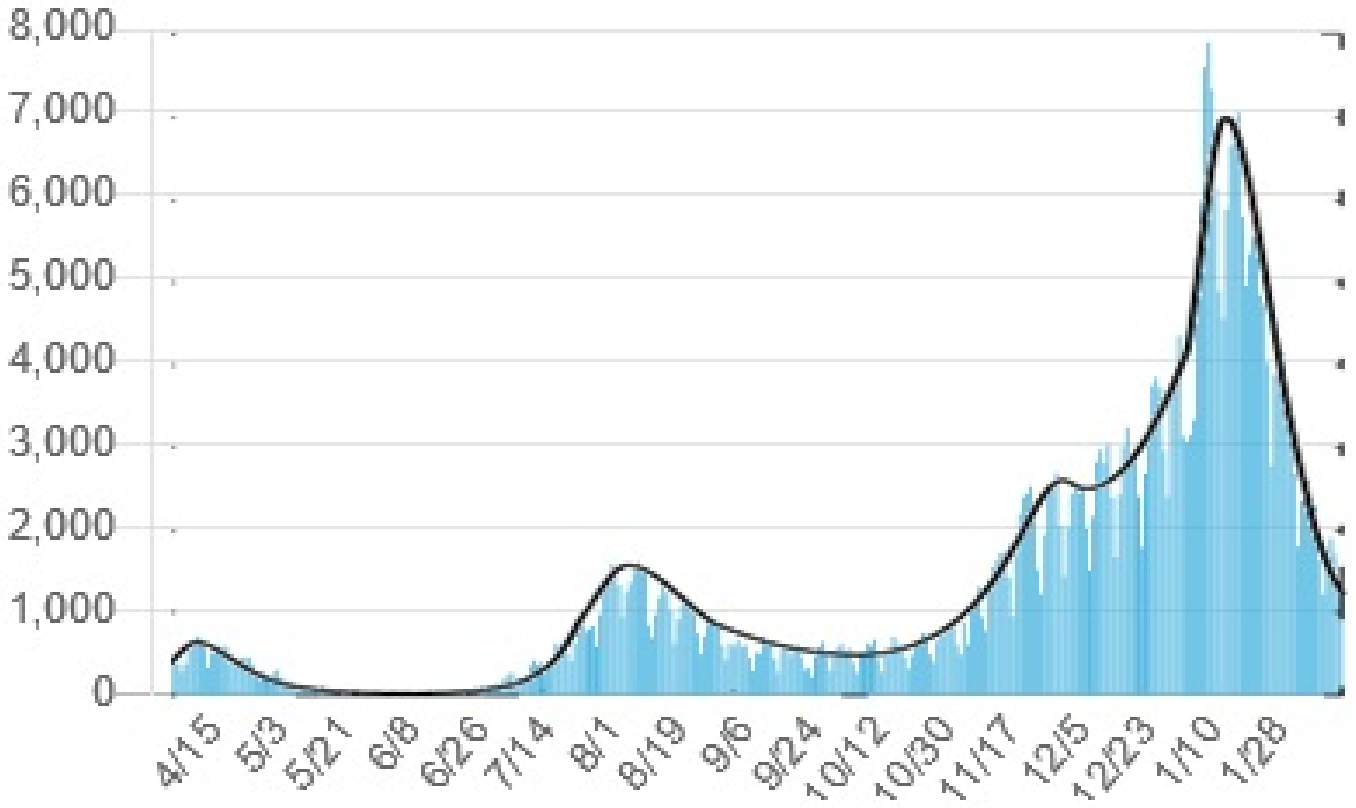}\hspace{1cm}
\includegraphics[width=5cm]{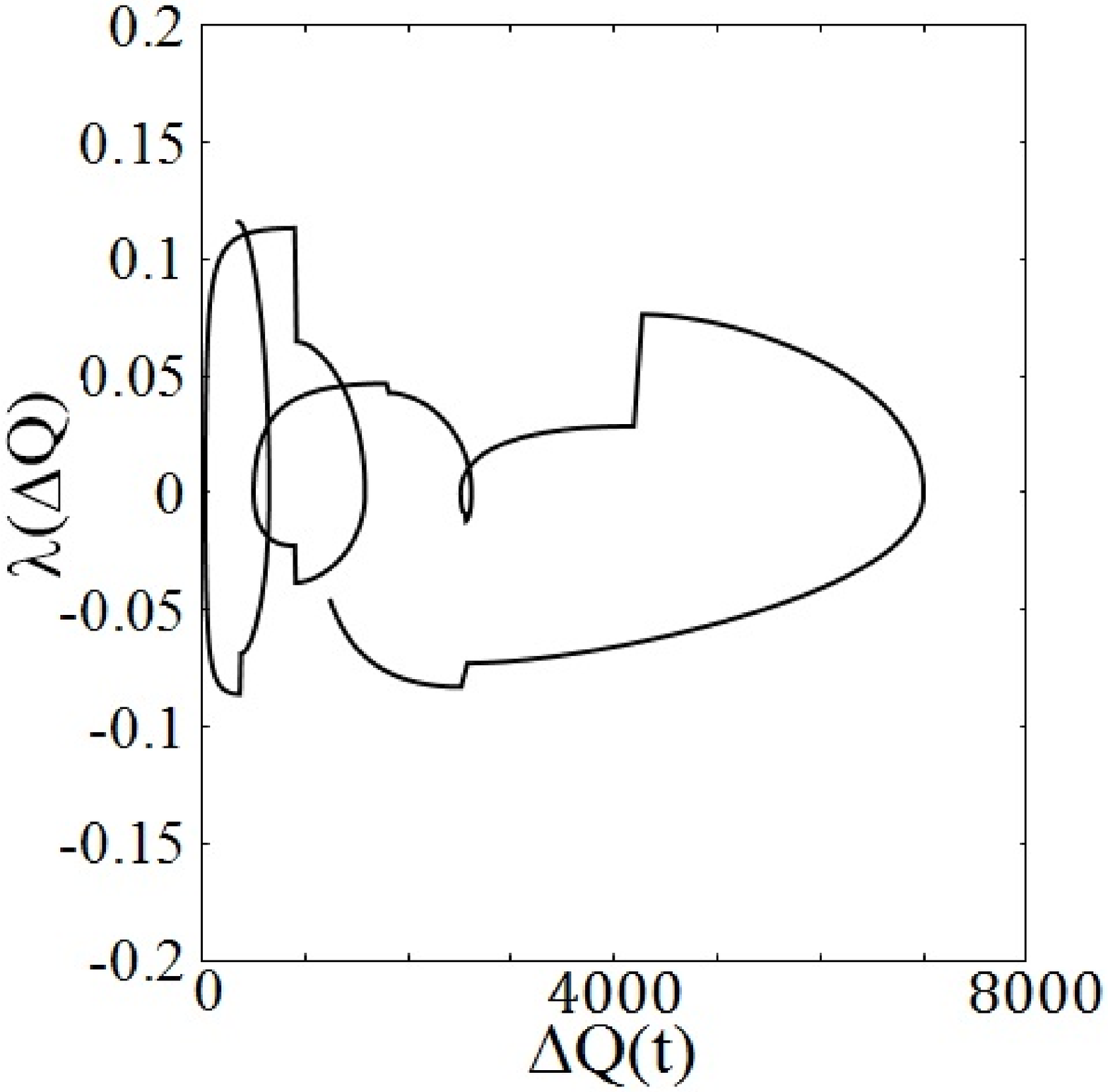}\\
(a)\hspace{5.5cm}(b)
\caption{\label{fig4}
(a) The time dependence of daily confirmed new cases in Japan
from April 5, 2020 to February 11, 2021. The dependence is fitted by piece-wise quadratic functions. 
(b) The net rate is shown as a function of the number of new cases.
The spiral nature of this plot indicates an enhancing wavy behavior of the infection curve.
The jumps seen in the plot are due to the procedure which does not impose the continuity
of the curvature.
}
\end{center}
\end{figure*}

Some of oscillations in biological systems such as the prey and predator system
have been explained by the Lotka-Volterra model \cite{Lotka, Volterra},
which is essentially a coupled logistic equation and it is reducible to a second order
non-linear differential equation for one variable.
Since the present model is based on a first order non-linear differential equation,
the origin of oscillatory solution of the present model is different
from that of the Lotka-Volterra model.

Several important implications of the present results are:\\
(1) In order to control the outbreak, a policy is needed to make $a > 1$
or $I_{\ell} < 0$ and $\lambda < 0$. Since $\lambda$ is determined
by $\beta$, $S$, $\gamma$ and $\alpha$(or $q$), this can be achieved by the lockdown measure
to reduce $\beta$, by the vaccination to reducing $S$ and
by the quarantine measure to increase $q$.\\
(2) The worst policy is $I_{\ell} > 0$. In this case, oscillation continues
until $\lambda$ becomes negative due to the herd immunity by vaccination and/or
infection of a significant fraction of the population.\\
(3) In order to make $\lambda$ negative, it has been rigorously shown that
increasing the quarantine rate $q$ is more efficient than reducing the
transmission coefficient $\beta$ by the lockdown measure \cite{oda-exact}.\\
This result indicates that the pandemic can be controlled only by keeping measures
of $\lambda < 0$ till $I =0$.\\
(4) It should be remarked that the change in the infectivity of the virus due to 
mutation can be included in $\lambda (I)$ in the present model.
Namely effects due to new variants of SARS-CoV-2 found in UK, in South Africa  or in Brazil
can be included by moving the state to a new $\lambda$ vs $I$ relation.

In this study, I assumed that $I_0$ is fixed and the dependence
of $\lambda$ on $I$ is symmetric. It is easy to generalize the present formalism
to the case of non-symmetric dependence of $\lambda$ on $I$.

\vspace{0.5cm}
\noindent
{\large\bf Acknowledgments}\\
This work was supported in part by JSPS KAKENHI Grant Number 
18K03573.


\begin{thebibliography}{99}
\bibitem{JHU}
Coronavirus Resource Center, Johns Hopkins University\\
https://coronavirus.jhu.edu/
\label{JHU}

\bibitem{oda-suda}
T. Odagaki and R. Suda,
https://doi.org/10.1101/2020.12.17.20248445
\label{oda-suda}

\bibitem{hethcote}
H. W. Hethcote,
SIAM Rev. {\bf 42}, 599-653 (2000).\\
http://doi:10.1137/S0036144500371907
\label{hethcote}

\bibitem{earn}
D. J. D. Earn,
Mathematical epidemiology. Lecture Notes in
Mathematics, vol. 1945 (eds. F. Brauer, P. van den Driessche, and J. Wu)
3-17 (Springer, Berlin, Germany, 2008).\\
https://doi.org/10.1007/978-3-540-78911-6
\label{earn}

\bibitem{zhang}
X. Zhang, C. Shan, Z. Jin and H. Zhu,
J. Differ. Equ. {\bf 266}, 803-832 (2019).\\
https://doi:10.1016/j.jde.2018.07.054
\label{zhang}

\bibitem{greer}
M. Greer, R. Saha, A.Gogliettino, C. Yu and K. Zollo-Venecek. 
R. Soc. open sci. {\bf 7}, 191187 (2020).\\
https://dx.doi.org/10.1098/rsos.191187
\label{greer}

\bibitem{oda-wavy}
T. Odagaki,
Sci. Rep. {\bf 11}, 1936 (2021). \\
https://doi.org/10.1038/s41598-021-81521-z
\label{oda-wavy}

\bibitem{SIR}
W. O. Kermack and A. G. McKendrick,
Proc. Roy. Soc. A {\bf 115}, 700-721 (1927).\\
https://doi.org/10.1098/rspa.1932.0171
\label{SIR}

\bibitem{SEIR}
R. M. Anderson and R. M. May,
Science {\bf 215}, 1053-1060 (1982).\\
DOI: 10.1126/science.7063839 
\label{SEIR}

\bibitem{Hethcote-SIQR}
H. Hethcote, M. Zhien and L. Shengbing,
Math. Biosciences {\bf 180}, 141-160 (2002).\\
https://doi.org/10.1016/S0025-5564(02)00111-6
\label{Hethcote-SIQR}

\bibitem{oda-idm}
T. Odagaki,
Infect. Dis. Model. {\bf 5}, 691-698 (2020).\\
https://doi:10.1016/j.idm.2020.08.013
\label{oda-idm}

\bibitem{Lotka}
A. J. Lotka,
Proc. Natl. Acad. Sci. {\bf 6}, 410-415 (1920).\\
https://doi.org/10.1073/pnas.6.7.410 
\label{Lotka}

\bibitem{Volterra}
V. Volterra,
Proc. Edin. Math. Soc. {\bf 6}, 4-10 (1939).\\
https://doi.org/10.1017/S0013091500008476
\label{Volterra}

\bibitem{oda-exact}
T. Odagaki,
Physica A{\bf 564}, 125564--1-9 (2021).\\
https://doi.org/10.1016/j.physa.2020.125564
\label{oda-exact}

\end{thebibliography}
\end{document}